\documentstyle[epsfig]{aipproc}

\def\www#1{{\tt #1}}

\begin{document}
\title{In ice radio detection of GZK neutrinos}

\author{David Seckel}
\address{Bartol Research Institute\\
University of Delaware\\
Newark Delaware 19716}

\maketitle

\begin{abstract}
Models for the source and propagation of cosmic rays are stressed by 
observations of cosmic rays with energies $E>10^{20}$ eV. 
A key discriminant between different models may be complementary 
observations of neutrinos with energies $E>10^{18}$ eV. Independent
of the source of the cosmic rays, neutrinos are produced during 
propagation via the GZK mechanism. Event rates for GZK neutrinos are 
expected to be in the range of $0.01-0.1$ per km$^3$ yr, suggesting a 
detector mass in excess of 1 Eg. Detection of radio cherenkov emission from showers 
produced in Antarctic ice may be an economical way to 
instrument such a large mass. It is suggested that a 100 km$^2$ array of 
antennas centered on Icecube may allow confirmation of the radio 
technique and also increase the science achievable with Icecube by 
providing vertex information for events with throughgoing muons.

\end{abstract}

\section*{Introduction}

Observations of cosmic rays with energies greater than $10^{20}$ eV 
present a problem. Assuming the particles are baryons, energy losses due 
to pair production and photoproduction demand that the sources of such 
particles be located within about 10 Mpc. Since the Universe is fairly 
lumpy on those length scales, it is somewhat surprising that the events 
appear to be distributed smoothly on the sky. Further, there are no 
obvious candidate sources when you look back along the arrival 
directions. Other particle choices present no better  
alternatives. Faced with this conundrum, source models generally fall into 
two classes. a) Astrophysical sources which satisfy the locality and 
isotropy constraints via a set of just so conditions. b) Models motivated 
by particle physics, which invoke the injection of high energy particles 
through the decay of massive particles or topological defects. 
These models may be distinguished through their associated 
neutrino fluxes.  

Most astrophysical cosmic ray acceleration models involve proton 
acceleration in a modestly dense environment. Within that source region, 
interactions may take place of the form $p+X\rightarrow n+\pi^+ + X'$, 
where $X$ and $X'$ may be anything. Two things happen at this juncture. 
1) The neutron may escape from the acceleration region. 2) The $\pi^+$ 
decays, ultimately producing an $e^+$, $\nu_e$, $\nu_\mu$ and an 
$\bar\nu_\mu$. Such models produce neutrinos at energy $E$ in comparable 
abundance to nucleons at energy $10 E$. It is also possible to have 
acceleration in low density environments, with no associated neutrino 
production.  
By contrast, the particle physics models result in the production 
of quark jets, which fragment into mesons, and eventually produce 
neutrinos, other leptons or 
photons. Baryon production in the fragmentation process accounts for a 
few percent of all particles, so the neutrinos outnumber the baryons.
Finally, once energetic protons are released into the cosmological medium, 
they produce more neutrinos through the GZK 
process (Griessen, Zatsepin, Kuzmin) involving photoproduction of pions 
from collisions with microwave background photons. 
Since protons with energies above the GZK theshold have been  
observed, one cannot avoid the conclusion that neutrinos are 
produced during cosmic ray propagation. 

With these comments, one can distinguish source models by their neutrino 
fluxes. An absolute minimum flux is the GZK flux inferred from the cosmic 
ray observations themselves. If this is all, then one is inclined to 
astrophysical acceleration models where the acceleration takes place is 
low density environments, such as at the interface of galactic winds, the 
termination shocks of jets from active AGN, etc. If the neutrino fluxes are 
modestly in excess of the predicted GZK fluxes, then acceleration in a 
dense astrophysical environment is favored, such as within AGN jets, 
around magnetars, or in gamma ray bursts, etc. Finally, if neutrino 
fluxes are greatly in excess of GZK models, then exotic particle physics 
models are indicated. 

To understand the 
observations of high energy cosmic rays, it is imperative to make 
complementary observations of ultra high energy cosmic neutrinos. To 
ensure that one has reached a correct interpretation, one should be 
prepared to search for fluxes as low as the GZK fluxes. The rest of this 
contribution is devoted to estimating the size detector required  
and comparing the potential of proposed detection techniques. In doing 
this analysis, a rate of 100 events per yr is suggested to go 
beyond simple discovery level science, an argument which should 
be familiar to supporters of the AUGER project. Since total fluxes are at 
a minimum in astrophysical models, I focus on this possibility.

\section*{Event Rates} Event rates are estimated by convolving 
cross-section and neutrino flux. Calculations of the GZK neutrino flux 
are illustrated by Figure 1, based primarily on results from Yoshida and 
Teshima\cite{YoshidaT} (YT). They start with a simple cosmic ray injection 
model, a homogeneous distribution of sources with proton spectrum $dN/dE = 
P(t) E^{-2}$. The power law is motivated by shock acceleration models. 
Their source evolution is described by $P(t) = P_0 (1+z)^m$ for 
$z<z_{max}$ and zero otherwise, with $P_0$, $m$ and $z_{max}$ taken as 
parameters, and $(1+z)(t)$ described by the cosmological expansion. YT 
normalize their models by integrating the predicted cosmic ray flux with 
energies $E>10^{19.5}$ eV and comparing to the AGASA data. Since 
photoproduction is efficient at high energies, the normalization only 
counts protons that originate within a few tens of Mpc. It 
therefore fixes $P_0$, but is independent of the evolution model. YT 
models are labeled by ($z_{max}$, $m$).

\begin{figure}
\centerline{\epsfig{file=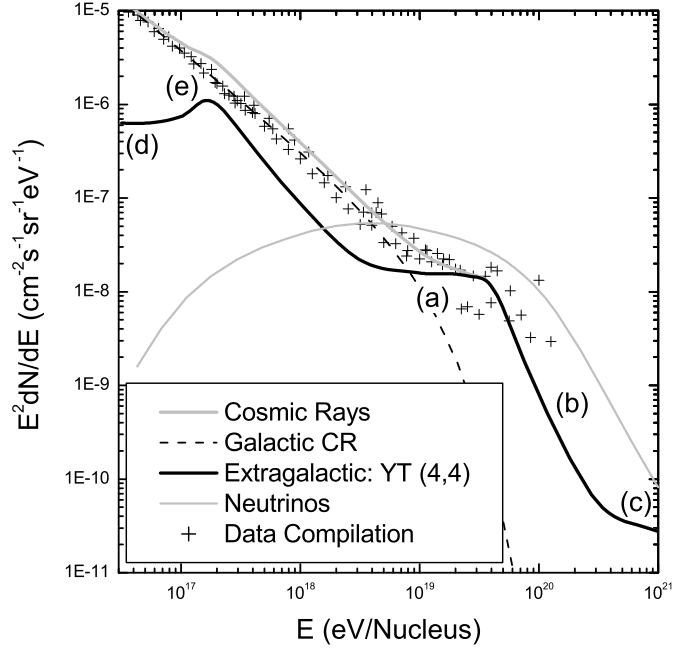,width=3.5in}}
\vspace{10pt}
\caption{Neutrino and cosmic ray flux for model (4,4) of Yoshida and 
Teshima.}
\label{fig1}
\end{figure}

In a homogeneous source model the GZK neutrinos are primarily produced at 
high red shift, so their flux depends both on $P_0$ and the source 
evolution. It follows that models with strong source evolution will 
produce more GZK neutrinos. In this regard, the model picked out for 
Figure 1 is the YT (4,4) model. This is the strongest source evolution 
model that YT consider and it gives a GZK neutrino flux approximately an 
order of magnitude stronger than their middle of the road (2,2) model. Even 
so it is interesting for several reasons. 

First, the evolved cosmic ray flux shows clearly several generic 
features. (a) The shelf at $10^{19}$ eV is fixed by the normalization and 
reflects the current injection rate integrated over the age of the 
universe. It does not, qualitatively, include injection at high redshift 
($z>1$) and so is relatively insensitive to parameters other than $P_0$. 
It is also relatively insensitive to the assumption of homogeneity unless 
magnetic diffusion effects are strong and sources are long lived. (b) The 
roll off above $10^{19}$ eV is due to $dE/dX$ from photoproduction. (c) 
The shelf at $10^{21}$ eV reflects the current injection rate integrated 
over the energy loss time. The level of (c) relative to (a) is sensitive 
to the assumption that the sources are distributed homogeneously. (d) The 
shelf below $10^{17}$ eV is due to injection at high red shift of protons 
at low enough energy that the only energy losses are adiabatic. The level 
of (d) relative to (a) is sensitive to the evolution parameters. 
(e) The bump at $10^{17}$ eV contains protons injected with high energy 
at high redshift that quickly loose energy due to photoproduction 
followed by pair-production. This bump is directly related to the flux 
level of GZK neutrinos. Its height is sensitive to $m$ and its position 
to $z_{max}$. 

Second, since publication of YT, it has become apparent that stronger 
source models are favored. For example, Engel and Stanev\cite{EngelS} argue 
for source evolution at least as strong as a (2,3) model, and they find 
conservative neutrino fluxes comparable to those from the YT (4,4) 
model. ES use a more detailed treatment of photoproduction than YT which 
may alter the evolution somewhat. They also use a different 
normalization, based on work by Waxman\cite{Waxman}, which involves 
integrating the cosmic ray flux above $10^{19}$ eV. Comparing YT's 
integral spectra to Waxman's suggests normalization differences of $\pm 
20\%$. Taken together this suggests the YT flux in Figure 1 may not be 
unreasonable for homogeneous models. 

Third, Figure 1 shows that when an extragalactic flux is combined with a 
reasonable galactic flux model interesting features relevant for 
consideration of GZK neutrinos may arise, beyond just the presence of a 
GZK cutoff. Specifically, in models of strong source evolution, the bump 
(e) should be noticeable as a distortion of the cosmic ray spectrum or 
its composition around $10^{17}$ eV. There is no hint of either effect in 
reviews by Gaisser\cite{Gaisser}.


The second half of the event rate calculation is the $\nu N$ cross 
section. Within the standard model, cross-sections at energies 
up to $10^{20}$ eV have been calculated by several 
authors\cite{GandhiQRS96,GandhiQRS98,GluckKR,KwiecinskiMS}. All such 
calculations are extrapolations, and should be treated with some 
caution. Typical variation in the charged current cross-section at 
$10^{19}$ eV is 20\% with larger variations in the neutral 
current cross-sections. Here, we use the results of Gl\"uck, 
Kretzer and Reya\cite{GluckKR} (GKR), which are a bit above average for 
charged currents and a bit below for neutral currents. 

Figure 2a shows event rates for different reactions assuming the ES flux 
and the GKR cross-sections. Rates assume $2\pi$ angular acceptance for 
downward going neutrinos, the upward flux being absorbed by the Earth, 
and are given in terms of neutrino energy, not shower energy. For charged 
current events total energy may be measurable based on known showering 
and $dE/dX$ properties of various leptons. Even $\tau$ leptons, at an 
EeV, lose some 7\% of their energy per km of path length in 
ice\cite{DuttaRSS}. For neutral current events only about 20\% of the 
primary neutrino energy\cite{GandhiQRS96} goes into the hadronic recoil, 
so event rates should be estimated using a neutrino threshold energy 
$\sim 5$ times the detector threshold. 

\begin{figure}
\centerline{\epsfig{file=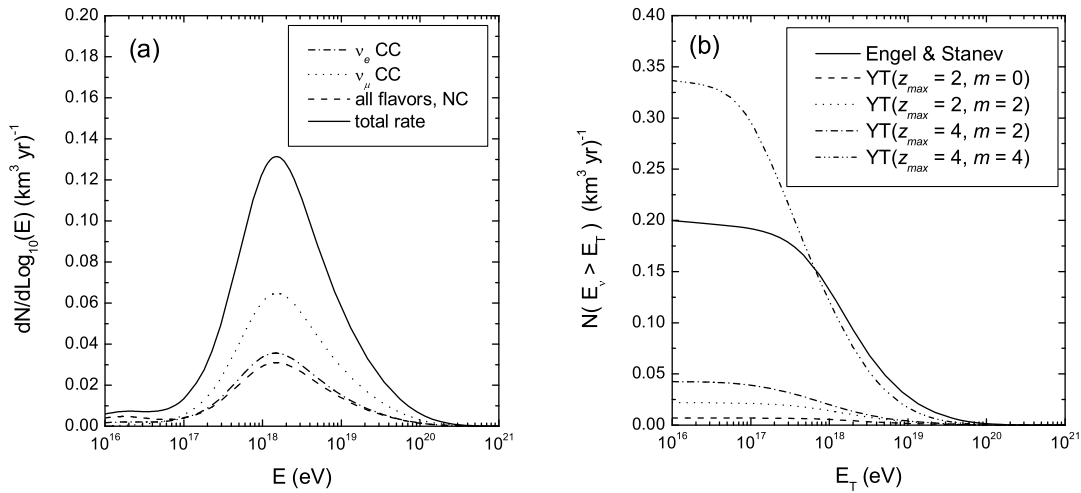,width=6. in}}
\caption{(a) Differential GZK neutrino interaction rates in ice, by 
flavor and event type, for the flux of Engel and Stanev. Rates 
for each flavor include both neutrinos and antineutrinos. The figure 
assumes no oscillations. (b) Integrated GZK neutrino event rates as a 
function of detector threshold. It is assumed that all energy is visible 
for charged current events and 20\% is visible for neutral current 
events.}
\vspace{-8pt}

\label{fig2}
\end{figure}

Figure 2b shows integral event rates for four YT models, as well as the 
ES flux. To achieve 100 events per yr requires roughly 1000 km$^3$ 
instrumented volume (or 1 Eg of mass) with threshold of 1 EeV. 
Pessimistic models, such as YT (2,2) may require 10 times the mass, but 
even then discovery could be accomplished on a modest 100 km$^3$ detector 
within a year or two of running. 

\section*{RICE and other technologies}

A summary of existing and proposed neutrino detectors is shown in Figure 
3. The most widely deployed technique is to use optical Cherenkov photons 
to reconstruct the tracks of muons produced in charged current reactions. 
In addition to muons, showers produced by the hadronic recoil in $\nu N$
scattering or by charged current electrons will produce 
intense concentrations of Cherenkov photons. Optical Cherenkov 
detectors are capable of energy determination for contained events and an 
extended sensitive volume for throughgoing heavy leptons. In the latter 
case, the energy resolution for the initial neutrino is likely to be 
quite poor. The largest optical Cherenkov detector constructed to date is 
AMANDA\cite{amandaweb}. Cubic kilometer detectors are planned, such as 
Icecube, but even these are too small to adequately probe GZK neutrinos. 
In principle, the effective volume for throughgoing muons may exceed that 
for contained events by an order of magnitude; however, without energy 
resolution the technique at best can set upper limits on EeV neutrino 
fluxes, but claims of GZK neutrino detection will be problematic.

\begin{figure}
\centerline{\epsfig{file=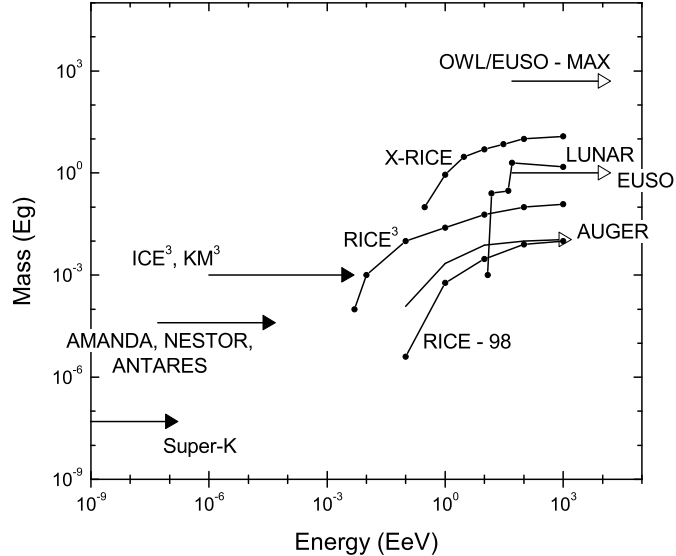,width=3.5 in}}
\vspace{15pt}
\caption{Effective mass of experiments proposed for detection of 
high energy neutrinos, as a function of neutrino energy. Experiments based 
on optical Cherenkov in water or ice are labeled by filled arrowheads. 
Open arrowheads denote air shower techniques. Radio Cherenkov experiments 
have dots overlayed and no arrowhead.}
\vspace{-10pt}
\label{fig3}
\end{figure}

Air shower techniques have also been discussed for neutrino detection. 
Soon, the largest air shower array will be the AUGER\cite{augerweb} experiment including 
some 3000 km$^2$ active area. However, since the column density of air is 
about 10 m, the instrumented mass corresponds to only about 0.03 Eg. The 
mass shown in Figure 3 accounts for the efficiency to detect neutrino 
induced cascades via the ground array\cite{augerweb}. Air showers may also 
be detected via flourescence techniques. To increase area, 
OWL\cite{owlweb}/EUSO\cite{eusoweb} have proposed to look down on $\sim 
1000 \times 1000$ km$^2$ of atmosphere from Earth orbit. The effective 
mass shown in the figure includes a 0.1 efficiency factor to include duty 
cycle and constraints on interaction depth and zenith angle so that the 
air shower can both fully develop and be cleanly separated from cosmic 
ray induced air showers. With these factors, space based flourescence 
techniques can provide 1 Eg of detector mass. However, if the indicated 
threshold of $5 \times 10^{19}$ eV\cite{eusoweb} is sharp, then using the 
ES flux would result in just 1 $\nu_e$ charged current event per yr, and less 
than one of other event types assuming that only the hadronic recoil is 
visible. For amusement, ``OWL/EUSO Max" refers to the total mass of the 
atmosphere utilized with 10\% efficiency.

A third option is to detect neutrinos through the coherent radio 
Cherenkov emission from the showers produced by neutrino interactions. 
Hopefully, it is not necessary, at this conference, to summarize the 
basic ideas behind the radio detection of high energy particles. 
One radio based technique involves the search for radio flashes from events in the lunar 
crust\cite{GorhamLN}. The effective mass shown in the figure is due to 
Alvarez-M\"uniz\cite{Alvarez}. It 
falls somewhat short of the mass and energy thresholds required for GZK 
neutrino detection. Additionally, only those events that occur 
well away from the limb of the moon will be separable from more frequent 
cosmic ray events causing similar signals.

The figure also shows sensitivity for experiments based on the 
technique of  deploying radio antennas in ice at the South 
Pole\cite{FrichterRM} (RICE). RICE-98\cite{riceweb} shows the effective 
mass of the current pioneering effort as configured at the end of the  98 
polar season.  RICE$^3$ depicts an array as may be deployed with 
Icecube. The lower threshold for RICE$^3$ arises from a relatively dense 
spacing of over 300 radio antennas. The largest version, X-RICE, is a 
10$^4$ km$^2$ array suggested for the detection of GZK neutrinos. The 
result shown is for antennas on a rectangular grid of 1 km 
spacing\cite{SeckelF}. 1 Eg of effective mass is achieved for $E> 1$ EeV.
Increasing the antenna density tenfold lowers the threshold, and 
allows a 1 Eg detector from 1000 km$^2$ of ice. (This is about 50\% 
efficiency since the ice is 2 km thick.) Allowing for detection of only 
electromagnetic and hadronic showers, such a detector would obtain 
approximately 30 $\nu_e$ and 25$\nu_\mu$ charged current, and 15 neutral 
current events from the Engel and Stanev GZK flux.

To summarize, optical Cherenkov experiments are too small, or too 
expensive, to ensure detection of GZK neutrinos. Air shower techniques 
require $10^5$ km$^2$ for particle detectors or $10^6$ km$^2$ for 
flourescence detectors due to the low density of air, beyond the scope of 
current ground based efforts. Space based flourescence experiments, as proposed, do 
not have a low enough threshold to detect GZK neutrinos. The lunar radio 
technique also has too high a threshold. Only the RICE technique seems to 
offer a suitable combination of mass and threshold to detect GZK 
neutrinos at the rate of $\sim 100$ neutrinos per yr.

\section*{Deployment strategies with Icecube}

The main difficulties with RICE are a) demonstrating that the physics of 
the technique is correctly described, b) establishing a calibration 
system for a deployed experiment, and c) overcoming the technical 
challenges of deploying on a remote basis up to 100 km from South Pole. 
(a) Recent experimental work by Saltzberg and 
collaborators\cite{Saltzberg}, and continuing theoretical efforts on 
several fronts\cite{BuniyR,AlvarezZ} suggests that the basic physics is 
sound, although there is still room for refinement. (b) One of the main 
points behind the RICE$^3$ plan is to provide explicit verification and 
calibration of a deployed experiment. (c) No serious work.

There are some flaws with the RICE$^3$ concept. First, Icecube is 
targeted toward detection of AGN neutrinos in the PeV region. Even with 
dense deployment, RICE$^3$ barely reaches below 10 PeV, although more 
sophisticated designs may improve on this situation. Second, the 
detectors do not really overlap. Optical transparency (bubbles) demands 
that Icecube be deployed in the bottom kilometer of ice, while radio 
transparency (geothermal warming) demands that RICE be deployed in the 
top kilometer. Since the Earth is opaque, most events in Icecube come 
from above or the side. The geometry of such events is such that the 
radio Cherenkov cone does not intersect a RICE array deployed above 
Icecube for most event vertecies contained in Icecube. The best geometry 
for coincident detection is charged current $\nu_\mu$ interactions where 
the muon passes through Icecube and RICE$^3$ gets the Cherenkov cone from 
the hadronic shower. For each zenith angle and impact parameter there is 
of order a linear km along the line of sight where the event geometry 
will satisfy this condition. Total effective overlap is about 1 km$^3$. 
Third, little progress is taken towards difficulty (c). Fourth, there 
may, in fact, be no AGN neutrinos, and fifth Icecube still cannot expect 
to see GZK neutrinos other than straggler external muons. 

An alternative deployment strategy is to make RICE$^3$ look like the 
central part of X-RICE, i.e. deploy $\sim 100$ antennas over 100 km$^2$ 
centered on Icecube. First off, such a deployment directly addresses 
difficulty (c). Second, such an array would provide vertex information 
for throughgoing muons detected by Icecube, allowing energy calibration 
of those events. Third, such an array would detect $\sim 10$ GZK neutrinos 
per yr, which would be a significant discovery. On the down side, the 
threshold for radio detection would rise, so potential overlap for 10 PeV 
AGN neutrinos is eliminated. Also, it seems that overlapping detection of 
GZK neutrinos still seems unlikely. Simple geometry considerations 
indicate that none of the ten GZK events are likely to be detected by 
Icecube. 


Which deployment strategy is better? In either case one cannot rely on 
the GZK neutrinos for calibration of RICE, even utilizing throughgoing 
muons. A bright AGN flux seems somewhat more likely than a bright flux in 
excess of GZK expectations at EeV energies. If neither bright option is 
present, then an independent confirmation and calibration of the radio 
technique is in order. Once achieved, the X-RICE prototype would likely 
give a first look at GZK neutrinos, while providing a technology 
development platform for the full X-RICE. On the other hand, the compact 
RICE$^3$ lacks the sensitive volume for GZK detection, and would 
leave unsolved the daunting technical problems of a large area Antarctic 
deployment.

I thank R. Engel for discussion and for sharing results of the 
ES model.

\end{document}